\documentclass[twocolumn,prd,showpacs,amsmath,amssymb,aps]{revtex4}

\usepackage{latexsym,amsmath,amssymb,color,mathrsfs}

\newcommand{\be}{\begin{eqnarray}}
\newcommand{\ee}{\end{eqnarray}}
\newcommand{\rar}{\rightarrow}
\newcommand{\Rar}{\Rightarrow}

\def\slash{ {\not} }
\def\dd{ \,\mathrm{d} }
\def\la{\langle}
\def\ra{\rangle}
\def\<{\langle}
\def\>{\rangle}
\newcommand{\exclude}[1]{}

\begin{document}

\title{The cosmological constant from the ghost. A toy model.}

\author{Federico R. Urban and Ariel R. Zhitnitsky}

\affiliation{Department of Physics \& Astronomy, University of British Columbia, Vancouver, B.C. V6T 1Z1, Canada}

\date{\today}

\begin{abstract}
We suggest that the solution to the cosmological vacuum energy puzzle is linked to the infrared sector of the effective theory of gravity interacting with standard model fields. We propose a specific solvable two dimensional model where our proposal can be explicitly tested. 
We analyse the 2d Schwinger model on a 2-torus and in curved 2d space, mostly exploiting the properties of its topological susceptibility, its links with the non-trivial topology or deviations from spacetime flatness, and its relations to the real 4d world.  The Kogut-Susskind ghost (which is a direct analogue of the Veneziano ghost in 4d) on a 2-torus and in curved 2d space plays a crucial r\^ole in the computation of the vacuum energy. The departure from Minkowski flatness, which is defined as the cosmological constant in our framework, is found to scale as $1/L$, where $L$ is the linear size of the torus. Therefore, in spite of the fact that the physical sector of 2d QED is represented by a single massive scalar particle, the deviation from Minkowski space is linear in $L$ rather than exponentially suppressed as one could na\"ively expect. 

 \end{abstract}

\maketitle

\section{Introduction}
This paper is the companion of our letter~\cite{4d} on the cosmological constant in four dimensions, where we compute the vacuum energy $\epsilon_{vac}$ in terms of QCD parameters and the Hubble constant $H$ with the result $\epsilon_{vac}  \sim H \cdot  m_q\la\bar{q}q\ra  /m_{\eta'} \sim (3.6\cdot 10^{-3} \text{eV})^4$, which is amazingly close to the observed value today $(2.3\cdot 10^{-3} \text{eV})^4$.

The Veneziano ghost in the 4d scenario bears close resemblance with the Kogut-Susskind ghost in 2d QED. The main goal of this paper is to test the idea that the vacuum energy esquires  $\sim 1/L$ correction to its Minkowski value,   with $L$ being  the linear size of the manifold.  This scaling is explicitely obtained in the exactly solvable case of the Schwinger model, thereby laying robust and firmly grounded basis for the results of~\cite{4d}. The key point   is that the corrections due to the very large but finite size $L$ of the manifold are small, $  \sim 1/L \sim H^{-1}$ but not exponentially small, $\exp (-L)$, as one could na\"ively anticipate for any QFT where all physical degrees of freedom are massive.
Such a scaling  is a result  of the Kogut-Susskind ghost in 2d QED and the Veneziano ghost in 4d QCD.  

{\bf A vacuum puzzle -} The Universe is accelerating away from us.  Or so is what a decade of experiments appeared to be at first suggesting, and now strongly confirming.  Indeed, since over ten years ago we have been accumulating experimental evidence supporting a non-zero cosmological energy density which appears to be non-clustered, homogeneously and isotropically distributed across the Universe~\cite{Spergel:2003cb,Riess:1998cb,Perlmutter:1998np} (see also~\cite{Copeland:2006wr,Sarkar:2007cx} for more up-to-date references).  Although there persists room for different explanations~\cite{Mustapha:1998jb,Celerier:1999hp,Tomita:2000jj,Zibin:2008vk}, and the observational data may need to be taken with some precautions~\cite{Sarkar:2007cx}, the so called ``Concordance Model'' (or $\Lambda$CDM for Cold Dark Matter), despite its disquieting implication that we do not know what the great majority of the Universe is made of, is nowadays widely accepted.

In numbers, what observational results tell us is that the Universe is permeated with an unknown form of energy density which makes up for about 75\% of the total energy density, which appears to be exactly the critical ratio for which the three-dimensional spatial curvature is zero, i.e.,
\be\label{lambda-term}
\Omega_\Lambda h^2  \approx 0.36 \, .
\ee

Explaining this number has proven to be an especially sturdy problem to attack, for particle theorists and cosmologists alike.  Indeed, in spite of the mass of models that have been thought out (see~\cite{Copeland:2006wr} for a comprehensive review), the picture is still blur, as most models encounter some sort of conceptual or observational obstacle.  It is customary to associate the ``dark'' energy density with vacuum fluctuations, whose energy density would be proportional to the fourth power of the cutoff scale, linked to the highest energy wave modes, at which the underlying theory breaks down.  If this argument were true, we would be faced with a disagreement between theory and observation varying between 40 to 120 orders of magnitude.  Clearly, this can not be, and more complex ideas must be probed.  Notice that most models in the literature adopt the same view, and therefore try to cancel or suppress short distance vacuum fluctations in one way or another~\cite{Copeland:2006wr}.

{\bf Gravity as an effective interaction -} The general framework into which this work falls is that of gravity as a low energy effective field theory, not as a truly fundamental interaction. In such a case, the corresponding gravitons should be treated as quasiparticles which do not feel all the microscopic degrees of freedom, but rather are sensitive to the ``relevant excitations'' only. We  note that such a viewpoint represents a  standard effective lagrangian approach in all other fields of physics such as condensed matter physics, atomic physics, molecular physics, particle physics. In particular, in condensed matter physics, a typical scale of the problem is  in the eV range, which has nothing to do with the electron mass  which is in the MeV, or the nuclei mass in the GeV. The relevant quasiparticles simply do not know about MeV or GeV scales as in the effective lagrangian approach those scales are effectively tuned away and never enter the system.

We should say that this philosophy is neither revolutionary nor new, rather, it has been discussed previously in the literature, see some relatively recent papers~\cite{Bjorken:2001pe,Schutzhold:2002pr,Klinkhamer:2007pe,Klinkhamer:2008nn,Thomas:2009uh} and references on previous works therein. If we accept the framework of the effective quantum field theory for gravity, than the basic problem of why the cosmological constant is 120 orders of magnitude smaller than its ``natural'' Plank scale $M_{Pl}^4$ is replaced by a fundamentally different problem: which is the relevant scale that enters the effective theory of gravitation? This effective scale obviously has nothing to do with the cutoff scale $\sim M_{Pl}$ which is typically associated with the highest energy ultraviolet (UV) scale, at which the underlying theory breaks down. Instead, the relevant effective scale must appear as a result of a subtraction at which some infrared (IR) scale enters the physics. What is important is that an effective quantum field theory (QFT) of gravitation has an IR parameter in its definition in contrast with the UV parameter which appears if gravitation is defined as a truly fundamental theory. 

According to this logic, it is quite natural to define the ``renormalised cosmological constant'' to be zero in Minkowski vacuum with metric $\eta_{\mu\nu}= \text{diag} (1,-1,-1,-1)$ wherein the Einstein equations are automatically satisfied as the Ricci tensor identically vanishes. Thus, the energy momentum tensor  $\langle T_{\mu\nu} \rangle \sim \eta_{\mu\nu}$ in combination with this ``bare cosmological constant'' must also vanish at the specific ``point of normalisation'' to satisfy the Einstein equations. Once this procedure is performed, the effective QFT of gravitation must predict the behaviour of the system in any non-trivial geometry of the space time. From this definition it is clear that all dimensional parameters, such as masses of particles and fields which contribute to the trace of the energy momentum tensor $\langle T_{\mu}^{\mu} \rangle$ in Minkowski vacuum, must cancel with the ``bare cosmological constant'' within an appropriate subtraction scheme, resulting in zero vacuum energy in Minkowski vacuum. This statement remains valid  for classically non-vanishing contributions to the trace of the energy momentum tensor $\langle T_{\mu}^{\mu} \rangle$ (due to massive particles) as well as quantum anomalous contribution such as nonzero gluon condensate in flat space.

This effect can therefore be understood as a Casimir type of vacuum energy.  Notice that the usual Casimir energies (e.g., from photons) are all typically irrelevant in understanding the observed vacuum energy, for they scale as $(L^2d^2)^{-1} \sim H^4$ where $d$ the distance between plates, $L$ is the size of the plates  and $H$ the Hubble parameter.

{\bf The shape of vacuum energy -} The arguments given above imply that a non-zero contribution to the energy density emerges only as a result of deviation from flatness, or, as we shall see below, from a spacetime with boundaries, and therefore must be proportional to some (positive) power of  $H$ in case of a de Sitter spacetime, or $1/L$ if we are dealing with a compact manifold of linear size $L$.  The chief question at this point is: which shape will this correction come in?  It has been known for a long time~\cite{Birrell:1982ix} that free massless particles contribute to the stress tensor through the conformal anomaly with a typical result
\be
\label{H^4}
\langle T_{\mu}^{\mu} \rangle \sim H^4 \, .
\ee
This is an astonishingly small number which can be ignored for all imaginable applications, as long as the curvature is small, which is the case today (see also~\cite{Thomas:2009uh}).

Contributions to the vacuum energy coming from, for example, scalar fields, have been calculated using a number of methods and renormalisation techniques, the most well known results being those coming in the shape of anomalous $H^4$ as mentioned above, or $M^2 H^2$ where $M$ is some grand unification mass~\cite{Sh.ch,Sola,Grande,Bauer,Ts1,Ts2}.  Most of these contributions are either hopelessly small or require physics beyond the Standard Model.  However, there is a combination whose vicinity to the observed value we think is worth exploring further; this combination reads
\be
\label{H}
\langle T_{\mu}^{\mu} \rangle \sim H \Lambda_{QCD}^3 \sim \left(10^{-3}  \text{eV}\right)^4 \, ,
\ee
instead of $\langle T_{\mu}^{\mu} \rangle \sim H^4$.  There are a number of arguments suggesting that the interactions can drastically change the na\"ive estimate (\ref{H^4}) especially if a non-local effective interaction (corresponding to an induced  long distance interaction) emerges. In that case, see refs.~\cite{Bjorken:2001pe,Schutzhold:2002pr,Klinkhamer:2007pe,Klinkhamer:2008nn} a vacuum energy of the form~(\ref{H}) may arise.  It is quite instructive that $ \Lambda_{\mathrm{QCD}} $ appears in the problem. Indeed, QCD is the only fundamental strongly-interacting QFT realised in nature. The electroweak  theory is actually weakly coupled, therefore, it is unlikely that the corresponding almost non-interacting heavy degrees of freedom contribute to the vacuum energy with the definition for it stated above.  Notice that covariance seem to require that, holding on to the de Sitter example for definiteness, only even powers of $H$ have right to enter the expression for the vacuum energy~\cite{Sh.sf}; this may however be not true~\cite{RW}.

{\bf This work -} The concrete model realising~(\ref{H}) has been proposed in the companion paper~\cite{4d}, where we have made the  crucial observations    that  the   Veneziano ghost~\cite{Ven} (see also~\cite{Shore07} for a review)  becomes the messenger through which the information stored at very large distances in our curved 4d spacetime can propagate and interact with microscopic particle physics, and viceversa.  However, in 4d most of the calculations can not be done explicitely, in both flat and curved spacetime, due to the intrinsic difficulties of QCD and strongly interacting fields in general.  The main focus of this paper is to turn the discussion to a simpler (and almost completely solvable) model, the Schwinger model in 2d~\cite{Schw}, where the analogue of the Veneziano ghost's pole is the well known Kogut-Susskind pole~\cite{KS} (which in fact was the starting point of Veneziano in his proposal), thus providing solid and reliable basis for the conclusions drawn in~\cite{4d}.

The paper is organised as follows.  First of all (section~\ref{bos}), a brief review of the Kogut-Susskind mechanism will be presented, including its generalisation to curved space, with an extended discussion on why na\"ive bosonisation fails when applied to curved space.  In the following section~\ref{ward}, the relevant Ward identities (WI) which provide the link between the ghost's propagator and the chiral condensate will be derived, and the effects of the non-zero quark masses will be explicitely computed, all in Minkowski space, making use of the topological susceptibility of the model.  Section~\ref{curv} moves on to analyse the Schwinger model and 2d QED on a 2-torus and in curved 2d spacetime, and shows where and how a linear term in $H$ (or $L^{-1}$, $L$ being the torus linear size) arises.  The final section is devoted to a short summary and some comments on the results obtained, including their translatability to the 4d real world.

\section{The Kogut-Susskind model}
\label{bos}

\subsection{Flat space}
The original KS model takes off from the 2d massive Schwinger lagrangian (that is, 2d QED with one massive fermion) given by
\be\label{fermiL}
{\cal L} = i \bar\psi \stackrel{\leftrightarrow}{\slash\partial} \psi - q \bar\psi \slash A \psi - \frac{1}{4} F_{ab} F^{ab} - m \bar\psi \psi + \mathrm{g. f.} \,\,\, ,
\ee
where g.f.\ stands for gauge-fixing terms, and the abelian field strength is given by
\be
F_{ab} = \partial_a A_b - \partial_b A_a \, ;
\ee
see the appendix for the other definitions and conventions.  This model is shown to be equivalent to a bosonic system whose lagrangian is
\be\label{KS}
{\cal L} &=& \frac{1}{2} \partial^a \hat\phi \partial_a \hat\phi + \frac{1}{2} \partial^a \phi_2 \partial_a  \phi_2 - \frac{1}{2} \partial^a \phi_1 \partial_a \phi_1 - \frac{1}{2} \frac{q^2}{\pi} \hat\phi^2 \nonumber\\
&&+ \frac{\alpha}{\beta^2} {\cal N} \cos\left[ \beta \left(\hat\phi + \phi_2 - \phi_1 \right) \right] \, ,
\ee
where $\cal N$ means normal ordering and the parameters $\alpha$ and $\beta$ (and $m$) are finite.  Notice that $\beta = 4\pi$ needs no renormalisation in 2d~\cite{KS,Coleman,Mandel,Naon}.

Working in the Lorentz gauge $A_a$ is divergence-free and can be expressed in the form $q A_a = \epsilon_{ab} \partial^b \varphi$, and it is related to the KS bosons as $\varphi = \sqrt\pi \left( \hat\phi - \phi_1 \right)$.  This relation will come in handy later when it will be used to calculate the topological susceptibility of this model.

The three bosons satisfy the commutation relations
\be\label{comm}
\left[ \hat\phi \, , \, \partial_t \hat\phi \right] &=& i \delta^2 \, ,\nonumber\\
\left[ \phi_1 \, , \, \partial_t \phi_1 \right] &=& - i \delta^2 \, ,\\
\left[ \phi_2 \, , \, \partial_t \phi_2 \right] &=& i \delta^2 \, ,\nonumber
\ee
from where we evince that $\phi_1$ is a massless ghost field, and its propagator will have a negative sign (in Minkowski space).  The masslessness of this ghost is also important in what follows.

The cosine interaction term includes vertices between the ghost and the other two scalar fields, but it can in fact be shown~\cite{KS} that, once appropriate auxiliary (Gupta-Bleuler) conditions on the physical Hilbert space are imposed, the unphysical degrees of freedom $\phi_1$ and $\phi_2$ drop out of every gauge invariant matrix element, leaving the theory well defined, i.e., unitary and without negative normed physical states, just as in the 4d Lorentz invariant quantisation of electromagnetism.  Specifically, this is achieved by demanding that the positive-frequency part of the free massless combination $(\phi_2 - \phi_1)$ annihilates the physical Hilbert space:
\be\label{gb}
(\phi_2 - \phi_1)^{(+)} \left|{\cal H}_{\mathrm{phys}}\right> = 0 \, .
\ee

It is important to notice that the KS ghost 
$\phi_1$ explicitly enters the expression for the topological susceptibility 
 (without its companion $\phi_2$ ). It  has important phenomenological consequences, as will be explained below.

\subsection{Curved space}
Now let us turn to a 2d curved spacetime.  The task is to find the boson equivalent lagrangian to a curved Schwinger model, whose lagrangian is going to be the covariantised version of (\ref{fermiL}).  Again we work in the covariant Lorentz gauge, for which $D^\mu A_\mu = 0$, which, in 2d, implies
\be
D^\mu A_\mu &=& \partial^\mu A_\mu - g^{\mu\nu} \Gamma^\lambda_{\mu\nu} A_\lambda = 0 \nonumber\\
&\Rar& q A_\mu = \sqrt{-g} \epsilon_{\mu\nu} \partial^\nu \varphi \, .
\ee
Here the longitudinal degree of freedom of $A_\mu$ is taken to vanish.  This means that
\be
F^{\mu\nu} F_{\mu\nu} = - \frac{2}{q^2} {\Box} \varphi {\Box} \varphi \, .
\ee
Notice that, just as it happened for the flat case, the Jacobian of the transformation is independent of the dynamical fields, and can therefore be absorbed into the normalisation constant, and doesn't appear in the effective Lagrangian.

Proceeding further, one can rescale the fermion field according to
\be
\left\{
\begin{array}{l}
\Omega^{1/2} \bar\psi = \bar\chi \\
\Omega^{1/2} \psi = \chi
\end{array} \right. \, ,
\ee
which transforms the kinetic term as $\sqrt g i \bar\psi \slash D \psi = i \bar\chi \gamma^a \partial_a \chi$.  The Jacobian in this case is given by the trace anomalous term for a fermion, and is, in 2d, proportional to the Ricci scalar $R$, and therefore, since we are considering the background fixed, it can again be absorbed in the normalisation constant.

In order to decouple the fermion one performs a chiral rotation defined by
\be
\left\{
\begin{array}{l}
\chi = e^{i \gamma^5 \varphi} \eta \\
\bar\chi = \bar\eta e^{i \gamma^5 \varphi}
\end{array} \right. \, ,
\ee
and, consequently, we obtain
\be
i \bar\chi \gamma^a \partial_a \chi &=& i \bar\eta \gamma^a \partial_a \eta - \bar\eta \gamma^a \gamma^5 \eta \partial_a \varphi \nonumber\\
&=& i \bar\eta \gamma^a \partial_a \eta + \bar\eta \gamma^a \eta \epsilon_{ab} \partial^b \varphi \, ,
\ee
from which we see that the second term on the right cancels the interaction term.  In this case the transformation has a non-trivial Jacobian, which in this case is just the same as in flat spacetime, and is therefore given by
\be\label{jac}
J &=& \exp \left\{ \frac{i}{2\pi} \int \dd^2x \varphi \bar{\bar{\Box}} \varphi \right\} \nonumber\\
&=& \exp \left\{ \frac{i}{2\pi} \int \dd^2x \sqrt{-g} \varphi {\Box} \varphi \right\} \, ,
\ee
leading to the effective Lagrangian
\be\label{fermi}
\sqrt{-g} {\cal L} = i \bar\eta \gamma^a \partial_a \eta &+& \sqrt{-g} \frac{1}{2} \varphi \left[ \frac{1}{q^2} {\Box} {\Box} + \frac{1}{\pi} {\Box} \right] \varphi \nonumber\\
&-& m \Omega \bar\eta e^{2 i \gamma^5 \varphi} \eta \, .
\ee

The free Fermi action, that is, when $m=0$, can be bosonised right away (at least as long as we are working in an infinite 2d spacetime where the gauge field has no harmonic components~\cite{SW}).  Once the boson identification has been performed, in order to be able to analyse the model along the lines of Kogut and Susskind we should now separate explicitely the degrees of freedom hidden in the higher derivatives in the Lagrangian.  We work with the effective free Lagrangian
\be
{\cal L} = -\frac{1}{2} \phi \Box \phi + \frac{1}{2} \tilde\varphi \left[ \frac{\pi}{q^2} \Box \Box +  \Box \right] \tilde\varphi \, ,
\ee
where we have rescaled the field $\varphi$ according to $\sqrt\pi \tilde\varphi = \varphi$.  From this curved spacetime Lagrangian we easily infer the form of the propagator for the higher derivative field $\tilde\varphi$, that is
\be
&\sqrt{-g} \left[ \frac{\pi}{q^2} \Box \Box +  \Box \right] \tilde\triangle_F = \delta^2(x) \, , & \nonumber\\
&\tilde\triangle_F = \lim_{\rho\rar0} \left[ \triangle_F \left( q/\sqrt\pi , x \right) - \triangle_F \left( \rho , x \right) \right] \, , &
\ee
which is the sum of a massive (mass squared $q^2/\pi$) scalar, and a massless ghost-like scalar.  We could then substitute the $\tilde\varphi$ field with these two fields obtaining (in the original KS notation where $\phi \rar \phi_2$, while the massive scalar is $\hat\phi$ and the ghost is $-\phi_1$)
\be\label{boseL}
{\cal L} = -\frac{1}{2} \hat\phi \Box \hat\phi - \frac{1}{2} \phi_2 \Box \phi_2 + \frac{1}{2} \phi_1 \Box \phi_1 - \frac{1}{2} \frac{q^2}{\pi} \hat\phi^2 \, ,
\ee
where everything is in curved space, and the partition function would be given by
\be
{\cal Z} &=& N \int [ {\cal D} \hat\phi ] [ {\cal D} \phi_2 ] [ {\cal D} \phi_1 ] \exp \left\{ i {\cal S} \right\} \, ,\\
{\cal S} &=& \int \dd^2x \sqrt{-g} {\cal L}\, .
\ee

The question now is what happens with the fermion mass term, which in the flat space bosonised sine-Gordon lagrangian becomes precisely the cosine-type interaction $\sim \cos\left[ \beta \left(\hat\phi + \phi_2 - \phi_1 \right) \right]$.  In flat spacetime this identification is obtained by expanding in a power series in the fermion mass and in the cosine interaction \emph{once the constants appearing in both lagrangians have been properly renormalised}.  Indeed, while the mass of the physical scalar $\hat\phi$ is determined independently from renormalisation, this is not so for the equivalence relating the strength of the cosine interaction in the boson lagrangian with the mass of the fermion in the original Schwinger lagrangian.  This is entirely due to the renormalisation properties of these 2d models~\cite{KS,Coleman,Mandel,Naon}.

In curved spacetime a very similar expansion can be performed, and one is tempted to na\"ively identify the coordinate dependent fermion mass term in (\ref{fermiL}) with an equivalent coordinate dependent interaction in the bosonic lagrangian, such as $k\Omega\cos\left[ \beta \left(\hat\phi + \phi_2 - \phi_1 \right) \right]$ with $k$ the usual renormalised sine-Gordon coupling constant~\cite{Eboli}.  However, such an identification would be wrong because it is immediate to see that the resulting action would lead to a non-conserved stress energy tensor, despite the initial (fermionic) system not exhibiting any such non-conservation.  To rephrase and summarise this last concept, it is essential to realise that if we identify the coefficient of the mass term in the fermi system eq.~(\ref{fermi}) with the corresponding one in the interacting sine-Gordon model, we would leave the $\Omega$ in front of the cosine: this is wrong as the corresponding energy momentum tensor is patently not conserved but instead $D_\mu T^\mu_\nu \propto \partial_\nu \ln\Omega$, in sharp contradiction with the initial system (the covariant generalisation of~(\ref{fermiL})), which has a perfectly well behaved stress tensor.  The solution to this apparent paradox is that in curved space or on a non-trivial manifold the coefficient of the cosine interaction is still a constant, but it is \emph{numerically different} from its Minkoski counterpart, as we will prove shortly.

This point is crucial in the discussion outlined here, because all of the non-trivial interaction between the KS ghost $\phi_1$ and the physical degree of freedom $\hat\phi$ stems from this term.  Without a precise knowledge of what happens in a general spacetime we are not able to make any definite computation on the impact of the chiral condensate at non-zero quark masses.  Notice further that this term is not responsible for the appearance of the condensate itself, which can be studied very rigorously in the Schwinger model even in curved space~\cite{SW}, but it is responsible for the coupling between the ghost's pole and the physical field, which is the essence of the Veneziano solution of the U(1) problem, and which is the most relevant contribution for us here.  Indeed, in the chiral limit, the effects of curved spacetime to the condensate appear only as corrections in powers of the curvature (that is, proportional to $H^2$ in 2d de Sitter space).  We will comment in details this result in the next sections.

Simple bosonisation in curved space therefore does not predict the value of the chiral condensate mass, and we must resort to different strategies to be able to extract the effects of the KS ghost.  This is the topic unravelled in the next section, where the answer will be inferred with the help of the topological susceptibility for the Schwinger model.

\section{Ward identities and the topological susceptibility}\label{ward}
The principal tool we will be adoperating in this section are the chiral anomalous WI for the 2d Schwinger model.  This is so because we need to find a way to overcome the difficulties encountered in bosonising the curved model.  The WI are a saviour in this case, because they allow one to link local observables such as the chiral condensate, to topological quantities, which are therefore independent on curvature.  In this section we will be working in flat 2d space, turning to curved space and/or non-trivial topologies, only in section~\ref{curv}.

The 2d WI can be derived in different ways, one of them being performing a chiral rotation $\psi \rar \exp(i\gamma_5\alpha) \psi$ with $\alpha(x)$ an infinitesimal (fictiously gauged) parameter, and demanding that the generating functional for the connected part of the action, customarily defined as $W$, does not change.  For the derivation see the appendix; here we are interested only in the final form the WI takes, which is 
\be\label{WI}
\frac{i}{4} \int \dd^2x \left< T_W Q(x) Q(0) \right> = m \left<\bar\psi \psi \right>_{\theta=0} + {\cal O}(m^2) \, ,
\ee
where we have defined
\be\label{charge}
Q = \frac{q}{2\pi} \epsilon_{ab} F^{ab} \, ,
\ee
as the topological charge density.  In this WI we have explicitely appended a subscript $W$ in the T-product, standing for Wick T-product which includes the contribution from unphysical states, as opposed to the usual Dyson T-product in which only physical states contribute.  The two definitions in coordinate space differ by a delta function, which is essential if one has to satisfy the WI (\ref{WI}), as will be shown explicitely in the chiral limit.  The saturation of the WI is also at the heart of the Veneziano proposal to solve the notorious U(1) problem of 4d QCD, i.e., the absence of a partially conserved (in the sense of PCAC algebra) ninth axial-vector current.

Notice that the WI in the functional form~(\ref{wiW}) is valid in flat and curved spacetimes alike, and only its explicit realisations will be slightly different.  The WI eq.~(\ref{WI}) provides a very powerful link between a topology-related quantity on the l.h.s. and the physical value of the (2d) quark condensate.  In order to exploit this relation we first define the topological susceptibility as a sort of variance for the topological charge
\be\label{topDef}
\chi(k^2) = \frac{i}{4} \int \dd^2x e^{i k x} \left< T_W Q(x) Q(0) \right> \, ,
\ee
which, according to the WI~(\ref{WI}) is related to the chiral condensate as $\chi(0) = m \left< \bar\psi \psi \right>$ at $\theta=0$.  Therefore, if we are able to compute the topological susceptibility of a system, we automatically know the value of the chiral condensate multiply the mass.  Notice that the mass coefficient of $\left< \bar\psi \psi \right>$ is exactly the coefficient in front of the cosine term in the bosonised equivalent of the model, which is the coefficient we are not able to explicitely calculate following the standard path of bosonisation.  The important point is that the WI hands in an entirely different way to compute the same coefficient, using a totally different approach, and that is much more easily generalised (although not fully solved) to curved and/or topological non-trivial spacetimes.

\subsection{Detour to 4d.}
Here we make a short review of the relevant 4d results. First of all, it is useful to notice that the topological charge density $Q$ is in fact a total divergence, and as such does not contribute to any matrix element in perturbation theory, which means that the solution to the $\eta'$ problem must be sough after in the non-perturbative sector of 4d QCD.  Indeed, in four dimensions,
\be
Q_{4d} = \partial_a K_{4d}^a = \frac{\alpha_s}{8\pi} G_{ab} \tilde G^{ab} \, ,
\ee
where $G_{ab} = \partial_a {\cal A}_b - \partial_b {\cal A}_a - i g [{\cal A}_a \, , \, {\cal A}_b]$ is the gluon field (denoted as ${\cal A}_a$) strength and $\tilde G$ its dual, and $\alpha_s$ the strong coupling constant.  The constant $g$ is the gluon self-coupling.  Note that with this definition of $Q_{4d}$ the Jacobian in the 4d equivalent variation (\ref{deltaL}) will bear an extra factor of 2.

One can explicitely write the Chern-Simmons gauge-variant current $K_{4d}^a$ as
\be
K_{4d}^a = \frac{\alpha_s}{4\pi} \epsilon^{abcd} \mathrm{tr} {\cal A}_b \left( G_{cd} - \frac{g}{3} \left[ {\cal A}_c \, , \, {\cal A}_d \right] \right) \, .
\ee
Despite its total divergence structure, the (euclidean) spacetime integral of $Q$ however needs not to vanish, and it is in fact related to the topological quantity $n$ called topological charge as
\be
\int \dd^4x Q_{4d} = n \in \mathbb{Z} \, ,
\ee
and it is different from zero for field configurations  (instantons) that become pure gauge at infinity.  Now, looking back at the WI (valid in exactly the same form in 4d) eq.~(\ref{WI}), it is immediate to evince that if we want the topological susceptibility $\chi(0)$ to be non-zero, we automatically need an unphysical massless pole in the zero momentum correlation function $\left< K_a K_b \right>$ in such a way that
\be
k_a k_b \left< K_{4d}^a K_{4d}^b \right>_{k=0} \neq 0 \;\; \Rar \;\; \chi(0) \neq 0 \, .
\ee
The Veneziano ghost is exactly this unphysical massless pole which is needed to saturate the WI, and it physically corresponds to the periodicity of the 4d QCD potential with respect to a generalised coordinate related to the $\theta$ angle~\cite{Diakonov}.

To conclude, it is essential, especially in view of the application of this mechanism to curved space and the problem of the vacuum energy, to understand that the solution to the U(1) problem is deeply rooted in the infrared sector of 4d QCD, and has nothing to do with the ultraviolet properties of the theory.  This fact will prove to be of fundamental importance when we will see that it is precisely in this sector that a non-vanishing and small positive vacuum energy density arises as soon as the spacetime is taken to be curved: a feature that is investigated for the real 4d world in our letter~\cite{4d}.

\subsection{Explicit calculation with the KS ghost \\ The chiral limit}
All of the preceding paragraphs discussion translates almost without changes to two dimensions, the main differences being the fact that 2d QED is abelian and that the chiral symmetry is not spontaneously broken by the condensate but by the anomaly associated with the axial current (just as in 4d QCD with only one flavour).  The main motivation for studying this 2d model lies on its beautiful analytical properties, for it allows for most of the calculations that in 4d present unsurmountable difficulties, to be done explicitly, and in a few lines.  Indeed, the Schwinger model proper (that is, with a massless fermion) can be solved exactly, which is equivalent to saying that the fermionic determinant is known explicitely.  These non-trivial features made the Schwinger model   an important playgound to test and develop most of the ideas that can be only approximately followed analytically in their 4d counterparts~\cite{Abdalla}.

Let us be more specific and define the equivalent 2d quantities.  Recall that
\be\label{Qdef}
Q = \partial^a K_a = \frac{q}{2\pi} \epsilon_{ab} F^{ab} = - \frac{q}{\pi} E \, ,
\ee
where $E$ is the electric field.  This definition leads to the identification of the 2d Chern-Simmons current as
\be\label{Kdef}
K_a = \frac{q}{\pi} \epsilon_{ab} A^b \, .
\ee
We choose again to work in the Lorentz gauge, which allows one to express the gauge potential $A_a$ as divergence of a scalar field $\varphi$.  Therefore, the topological charge density can be written as $Q = q/\pi E = \bar{\bar\Box} \varphi /\pi$.  Moreover, as it has been shown in section \ref{bos}, the field $\varphi$ is identified with two degrees of freedom, one of which is the sought after massless ghost's pole.  We will now proceed on to show how the KS construction satisfies explicitely the WI.

First of all, we need to compute $i\left< T_W E(x) E(y) \right>$ in coordinate space.  This quantity is well known for the Schwinger model (with a massless fermion), and has been calculated using instanton solutions on 2d euclidean compact spaces such as 2-sphere or 2-torus, whose appropriate infinite space limits have then been taken~\cite{Jay,Sachs91,Smilga}.  The (euclidean) result, which we want to reproduce using the KS mechanism, is, for zero fermion mass
\be\label{helvetica}
\left< T_W E(x_E) E(0) \right> = \delta^2(x_E) - \frac{\mu^2}{2\pi} K_0(\mu |x_E|) \, ,
\ee
where the subscript $E$ stands for euclidean, and $\mu^2 = q^2/\pi$.  The $K_0$ is the modified Bessel function of order 0, which describes massive scalars in 2d.  Notice the presence of the $\delta$ function in this expression: without it it would be impossible to satisfy the WI for vanishing quark mass, as its integral over 2d will never be zero.  This $\delta^2(x_E)$ is precisely the contribution of the ghost's states, which necessarily tells us that the T-product appearing in (\ref{helvetica}) must be intended as Wick T-product.

We want to demonstrate that the KS prescription is what is needed to obtain the result (\ref{helvetica}), which had been previously calculated exploiting the instanton solutions possessed by the Schwinger model~\cite{Jay,Sachs91,Smilga}.  This can be done explicitely using the splitting of the $\varphi$ field in massive and massless ghost's degrees of freedom as
\be\label{splitdef}
\varphi = \sqrt\pi \left( \hat\phi - \phi_1 \right) \, ,
\ee
which now can be used in the (euclidean) correlation function of $E$ to give
\be\label{Ecorr}
\left< T_W \!\!\!\right.\!\!\!&&\!\!\!\left.\!\!\! E(x_E) E(y_E) \right> = \nonumber\\
&=& \frac{\pi}{q^2} \int \frac{\dd^2p_E}{\left(2\pi\right)^2} p_E^4 e^{-i p_E (x_E-y_E)} \left[ - \frac{1}{p_E^2+\mu^2} + \frac{1}{p_E^2} \right] \nonumber\\
&=& \int \frac{\dd^2p_E}{\left(2\pi\right)^2} e^{-i p_E (x_E-y_E)} \left[ 1 - \frac{\mu^2}{p_E^2+\mu^2} \right] \nonumber\\
&=& \delta^2(x_E - y_E) - \frac{\mu^2}{2\pi} K_0(\mu |x_E - y_E|) \, ,
\ee
which is the result (\ref{helvetica}), as anticipated.  Integrating this equation with $y_E \rar 0$ and at zero momentum gives
\be
\int \dd^2x_E \left[ \delta^2(x_E) - \frac{\mu^2}{2\pi} K_0(\mu |x_E|) \right] = 0 \, .
\ee
Hence, the WI is satisfied when $m \rar 0$, \emph{owing to the presence of the ghost's gapless excitation.}

The calculation in Minkowski space goes on exactly in the same way, using (\ref{splitdef}):
\be\label{Mcorr}
i \left< T_W \!\!\!\right.\!\!\!&&\!\!\!\left.\!\!\! E(x) E(y) \right> = \frac{i}{q^2} \left< T_W \bar{\bar\Box}_x \varphi \bar{\bar\Box}_y \varphi \right> =\nonumber\\
&=& \frac{i\pi}{q^2} \int \frac{\dd^2p}{\left(2\pi\right)^2} p^4 e^{-i p (x-y)} \left[  \frac{i}{p^2-\mu^2} - \frac{i}{p^2} \right] \nonumber\\
&=& \int \frac{\dd^2p}{\left(2\pi\right)^2} e^{-i p (x-y)} \left[ - 1 - \frac{\mu^2}{p^2-\mu^2} \right] \nonumber\\
&=& - \delta^2(x - y) - i \frac{\mu^2}{2\pi} K_0\left( \mu \sqrt{-\left( x-y \right)^2} \right) \, .
\ee
Here the 2d massive scalar propagator is $\hat\Delta_F = (i/2\pi) K_0$, in coordinate space.  It is important to notice that had we used the equations of motion for $\varphi$, that is $\bar{\bar\Box} \varphi = - \hat\phi /\sqrt\pi$ (which corresponds to the Dyson T-product), we would have missed the important $\delta^2(x)$ factor, without which the WI would not be satisfied.

The next step is to calculate the momentum-space topological susceptibility as
\be\label{topflat}
\chi(k^2) &=& \frac{i}{4} \int \dd^2x e^{i k x} \left< T_W Q(x) Q(0) \right> = \nonumber\\
&=& i \left( \frac{q}{2\pi} \right)^2 \int \dd^2x e^{i k x} \left< T_W E(x) E(0) \right> \nonumber\\
&=& \left( \frac{q}{2\pi} \right)^2 \left( -1 - \mu^2 \hat\Delta_F(k^2) \right) \, ,
\ee
where $\hat\Delta_F(k^2) = (k^2 - \mu^2)^{-1}$ is the propagator for the physical massive scalar $\hat\phi$.  The topological susceptibility goes to zero as it should when $k^2 \rar 0$.  The physical reason behind this key result is that the ``would-be Nambu-Goldstone boson"   associated with the chiral U(1) symmetry cancels exactly with the massless ghost.  This cancellation is exact as long as the quark has zero mass, but a finite residue would remain otherwise, as we shall see in what follows.

\subsection{Explicit calculation with the KS ghost \\ Adding a small quark mass $m$}
In order to draw the parallel with the Veneziano ghost in 4d we need to investigate what happens when a non-zero mass term for the fermion is introduced.  In this case the bosonised flat space lagrangian contains the cosine interaction term, which introduces loop corrections in the expression for the topological susceptibility (\ref{topflat}).  These corrections can be found in complete analogy with the 4d Veneziano's computation by ``dressing'' the $\hat\phi$ propagator as
\be\label{propcorr}
\hat\Delta_F^I &=& \hat\Delta_F \left( 1 + \frac{m_0^2}{p^2-\mu^2} + \ldots \dots \right) \nonumber\\
&=&\frac{1}{p^2-\mu^2-m_0^2} \, ,
\ee
where $m_0^2 \simeq - m \left< \bar\psi \psi \right>$, $m$ being a small quark mass.  This dressed propagator is the source of the saturation of the WI at non-zero quark mass, since one immediately see that upon performing the 2d $\dd^2x$ integration to obtain $\chi$, the two contributions do not cancel but leave a finite, negative, remnant
\be\label{topflatM}
\chi(0) = \frac{1}{4}\mu^2 \left( -1 + \frac{\mu^2}{\mu^2 + m_0^2} \right) \simeq - m_0^2 \simeq m \left< \bar\psi \psi \right> \, ,
\ee
as it should, when the quark mass is small, and the WI~(\ref{WI}) is satisfied.  Let us stress once more that the WI is saturated only thanks to the ghost's contribution.  The ghost, as clearly stated in the original paper by Kogut and Susskind, plays no r\^ole when it comes to compute gauge invariant martix elements, because the Gupta-Bleuler conditions imposed on the physical Hilbert space make it decouple from these observables.  Moreover, in this way the theory is automatically unitary.  However, some quantities like the topological susceptibility do depend on the gauge-variant, and not observable, current $K_a$ in a non-trivial way, renewing the importance of the decoupled ghost's states even in physical observables, in this case the 2d analogue of the $\eta'$ mass.

It is of pivotal importance to realise that the arguments just laid are robust against perturbation theory.  Indeed one may worry that interactions not only would act on the physical massive pole, but could also shift the two massless poles $\phi_1$ and $\phi_2$ necessary for the realisation of the KS mechanism.  However, the KS dipole poles at zero mass stay there, and are thus ``protected''~\cite{Weinberg}.

In the forthcoming section we will repeat these steps for spacetimes with boundaries, e.g.\ a 2-torus, and outline the calculation for a general curved spacetime.  In the simple case of a torus with non-trivial boundary conditions, an explicit \emph{linear} dependence on the size of the manifold will appear.

\section{Topology and curvature}\label{curv}
Let us consider now the Schwinger model, and its generalisation allowing for a small quark mass, on a compact manifold.  In what follows we enclose the system in a box of length $L$, and, in order to include the effects of the spacetime curvature, we work with a general 2d metric; for simplicity, and to facilitate the comparison with the literature, the metric will have euclidean signature, see the appendix.

\subsection{Quantisation on a torus}
To begin with, we shall show that, in the chiral limit, the topological susceptibility for the compact 2d \emph{flat} spacetime is still zero, as imposed by the WI~(\ref{WI}).  To this end, we notice that the only difference in the previous section's calculation is the fact that the integrals run from 0 to $L$ and that we will need to use the discretised version of the (euclidean) scalar massive propagator, which is given by
\be\label{discrete}
\hat\Delta_F(x_E) &=& \\ =\frac{1}{L_1 L_2} \!\!\!\!\!&\!\!\!\!\!\!&\!\!\!\!\!\! \sum_{n1,n2} \frac{e^{2\pi i (n_1 x_E^1 / L_1 + n_2 x_E^2 / L_2)}}{\left( 2\pi / L_1 \right)^2 n_1^2 + \left( 2\pi / L_2 \right)^2 n_2^2 + \mu^2} \, . \nonumber
\ee
Now, if we insert this expression in the (euclidean and discretised version of) eq.~(\ref{topflat}), with the appropriate limits of integration, we again find that the topological susceptibility vanishes when $k^2\rar0$.  Indeed:
\be\label{checkTorus}
\int_0^L \dd^2x \sqrt{g_E} \left( \delta^2(x_E) - \mu^2 \hat\Delta_F(x_E) \right) = 1 - 1 = 0 \, ,
\ee
where $\sqrt{g_E} = \tau_0$ (see the appendix).  More detailed arguments supporting this form for the topological susceptibility will be given in the next subsection.

Precisely the same calculation can now be performed when the quark has a small but non-zero mass, as in (\ref{propcorr}), with the result
\be
\label{topflatT}
\chi(0)^{\mathrm{torus}} = - \frac{\tau_0}{4} \frac{\mu^2 m_0^2}{\mu^2 + m_0^2} \, ,
\ee
which is apparently exactly the same one quoted above, a part from the $\tau_0$ factor in front.  However, there is a very important subtlety that enters this expression (\ref{topflatT}), namely the fact that the value of $m_0^2$, which  is defined as $m_0^2=-m\left< \bar\psi \psi \right>$, is \emph{not} the same in a compact space and in the full Minkowski space as the chiral condensate is different in these two cases.

The magnitudes for the chiral condensates on the finite torus has been derived a while ago~\cite{SW}, and it reads
\be
\label{condTorus}
\left< \bar\psi \psi \right> &=& \frac{1}{L|\tau|} \exp\left\{ - \frac{\pi}{\mu L \tau_0} \coth \frac{\mu L \tau_0}{2 |\tau|}  \right\} \,  \nonumber \\
&+&\frac{1}{L|\tau|} \exp\left\{
F(\tau,L) - H(\tau,L) \right\} \, ,
\ee
where
\be
F(\tau,L) &=& \sum_{k>0} \left( \frac{1}{k} - \frac{1}{\sqrt{k^2+a^2}} \right) \, ; \nonumber\\
H(\tau,L) &=& \sum_{k>0} \frac{1}{\sqrt{k^2+a^2}} \left( \frac{1}{e^{- 2 \pi i z_+} - 1} + \frac{1}{e^{2 \pi i z_-} - 1} \right);\nonumber\\
z_\pm &=& \frac{1}{|\tau|^2}\left( n \tau_1 \pm i \tau_0 \sqrt{k^2+a^2} \right) \, ;\nonumber
\ee
and $a = \mu L |\tau| / 2\pi$.  If we take the limit for which $\mu L \gg 1$ then the above expression simplifies to
\be
\label{condTexpl}
\left< \bar\psi \psi \right> &\simeq& \frac{1}{L |\tau|} \exp\left\{ - \frac{\pi}{\mu L \tau_0} + \gamma \right\}  \nonumber\\
&+& \frac{1}{L |\tau|}\left\{ \ln\frac{\mu L |\tau|}{4\pi} + \frac{\pi}{\mu L |\tau|} + \ldots \right\} \nonumber\\
&=& \frac{\mu}{4\pi} e^\gamma \exp\left\{ \frac{\pi}{\mu L} \frac{\tau_0 - |\tau|}{\tau_0 |\tau|} \right\} \nonumber\\
&\simeq& \frac{\mu}{4\pi} e^\gamma \left[1 + \frac{\pi}{\mu L} \frac{\tau_0 - |\tau|}{\tau_0 |\tau|} \right]  \, .
\ee

Let us briefly notice that, strictly speaking, the chiral condensate in Euclidean spacetime should not be written as $\left< \bar\psi \psi \right>$, but rather $\left< \psi^\dagger P_+ \psi \right>$, in the notation of~\cite{SW}, which has opposite sign compared to its Minkowski counterpart (this is the reason behind the positive sign in~(\ref{condTexpl})).

Expression~(\ref{condTexpl}) shows clearly that, if one is to employ a non-trivial torus, then automatically a linear correction in the size $L$ will occur.  This expansion for the chiral condensate can be used in the previous eq.~(\ref{condTorus}), and then expanded at first order in $\mu L$ to give
\be\label{linear}
\chi(0)^{\mathrm{torus}} \simeq \chi(0) \tau_0 \left[ 1 + \frac{\mu^2}{\mu^2 + m_0^2} \frac{\pi}{\mu L} \frac{\tau_0 - |\tau|}{\tau_0 |\tau|} \right] \, .
\ee
This is the most important result of this paper, for it shows how the linear dependence on the size of the manifold arises.  Since the quark mass $m$ is small, this expression further simplifies to
\be\label{linear2}
\chi(0)^{\mathrm{torus}} \simeq \chi(0) \tau_0 \left[ 1 + \frac{\pi}{\mu L} \frac{\tau_0 - |\tau|}{\tau_0 |\tau|} \right] \, .
\ee
It is very important to notice that the linear correction obtained in equations~(\ref{linear}) and~(\ref{linear2}) depends crucially on the existence of both the ghost \emph{and} the non-masslessness of the quark, the latter being of fundamental importance in connecting the topological susceptibility just obtained with the vacuum energy of the system.

This result is especially important since, as already pointed out, it allows one to derive explicitely the exact coefficient appearing in front of the cosine interaction term, for any given topology (and curvature, see below), without having to rely on dubious and ambiguous series expansions in the mass-interaction term.  We are therefore able to obtain the exact bosonised version of the   QED$_2$ for a given spacetime.

A last note concerning the interpretation of the result~(\ref{linear2}) as a finite temperature effect.  Having allowed for the most general $\tau$ in our computations, one expects to be able to extract the (small) finite temperature temperature behaviour of the system.  This is indeed possible within this framework, upon substituting $\tau = i \beta / L$ where $\beta = 1/T$ is the inverse temperature.  In doing so we have to switch back to Minkowski spacetime, and the finite-$T$ condensate would look like
\be\label{condTemp}
\left< \bar\psi \psi \right> = - T \exp\left\{ -\frac{\pi}{\mu} T + F(\tau,L) \right\} \, ,
\ee
which, in our limit for which $T\rar0$, gives
\be\label{condZeroT}
\left< \bar\psi \psi \right> \rar - \frac{\mu}{4\pi} e^\gamma \, ,
\ee
without linear corrections (notice the overall minus sign), which would instead appear at high $T$.  This can be seen using the general $\tau$ expression eq.~(\ref{condTexpl}) as well, using the fact that $\tau = i\beta/L \Rar |\tau| = \tau_0$.

\subsection{The effect of curvature}
When curvature is introduced, very similar effects to those depicted for a non-trivial compact space appear.  In this section the spacetime will be taken to be infinite.

First of all, the correct definition of the topological susceptibility needs to be found.  In doing this, our guide will still be the WI~(\ref{WI}), which states that in the chiral limit the spacetime integral of the topological charge correlation function is zero.  The obvious generalisation to a general 2d curved space of this expression is
\be\label{topCurved}
\chi(k^2)^{\mathrm{curved}} = \frac{i}{4} \int \dd^2x \sqrt{-g} e^{i k x} \left< T_W \tilde Q(x) \tilde Q(0) \right> \, ,
\ee
where $\tilde Q(x)$ is the curved space topological charge density.  It is clear that, being entirely a topological quantity in nature, this charge density will roughly speaking behave as
\be\label{Qcurved}
\int \dd^2x \sqrt{-g} \tilde Q(x) = n \quad\Rar\quad \tilde Q(x) \propto Q(x) / \sqrt{-g} \, .
\ee

Notice that the topological susceptibility is not, strictly speaking, a topological invariant, as it depends on the curvature of the spacetime rather than solely on its topology, because the integration is only performed over $\dd^2x$ and not over $\dd^2x\dd^2y$.

To show that the definition~(\ref{topCurved}) for the topological susceptibility is consistent with the WI, we notice that the curved space propagators which will appear, see equation~(\ref{Ecorr}) or~(\ref{Mcorr}), are defined by the massive curved space Klein-Gordon equation
\be
\left( \Box + \mu^2 \right) \hat\Delta_F^{\mathrm{curved}}(x) = - \frac{\delta^2(x)}{\sqrt{-g}} \, ,
\ee
and therefore it is normalised, employing sensible boundary conditions on the derivatived of the Green's function, as
\be\label{norm}
\mu^2 \int \dd^2x \sqrt{-g} \hat\Delta_F^{\mathrm{curved}}(x) = -1 \, ,
\ee
which automatically ensures the validity of the WI~(\ref{WI}).  The WI must also be satisfied by directly plugging in the explicit expression for the propagator, which, for instance in 2d de Sitter spacetime, reads~\cite{Bunch}
\be\label{deSprop1}
\hat\Delta_F^{\mathrm{curved}}(x,y) = i \theta(t) G_+(x,y) - i \theta(-t) G^*_+(x,y) \, ,
\ee
where the two point function $G_+$ is given by
\be\label{deSprop2}
G_+(x,x_0) &=& \frac{1}{2\pi^2}\sqrt{\eta\eta_0} \int_{-\infty}^{+\infty} \!\!\!\!\dd k e^{i k x} K_\nu(ik\eta) K_\nu(-ik\eta_0) \nonumber\\
&=& \frac{1}{4\pi} \Gamma\left( \frac{1}{2} - \nu \right) \Gamma\left( \frac{1}{2} + \nu \right) \times\nonumber\\
\,\times\phantom{x}\!\!\!_2F_1\left[ \frac{1}{2} - \nu \!\!\!\!\!\!\right.&,&\left.\!\!\!\! \frac{1}{2} + \nu ; 1 ; 1+ \frac{(\eta - \eta_0)^2 - (x - x_0)^2}{4\eta\eta_0} \right] \, .
\ee
Here $K_\nu$ is the Bessel function $K$ of order $\nu$, where $\nu^2 = 1/4 - \mu^2 / H^2$ ($H$ is the Hubble parameter), while $_2F_1$ is an hypergeometric function.  The propagator is expressed in conformal coordinates
\be
\dd s^2 = \Omega^2 \left( \dd \eta^2 - \dd x^2 \right) \quad\mathrm{with}\quad \Omega^2 = 1 / H^2\eta^2 \, .
\ee

Obviously, in this case an explicit computation is much harder, but as the flat torus example showed us, we should expect the appearance of a non-zero topological susceptibility as the quark mass becomes non-vanishing, and this susceptibility is   proportional to $m_0^2$, since it must vanish when $m=0$.  Once more, we can borrow the results of~\cite{SW} to show that in this case (2d de Sitter spacetime) the first correction to the condensate is given by
\be\label{condCurv}
\left< \bar\psi \psi \right>^{\mathrm{curved}} = - \frac{\mu}{4\pi} e^\gamma \, e^{-H^2/6\mu^2} \,  .
\ee
For sufficiently small quark mass $m$ the topological susceptibility behaves as
\be\label{topCurv}
\chi(0)^{\mathrm{curved}} &=&m\left< \bar\psi \psi \right>^{\mathrm{curved}} \nonumber\\
&\simeq& - m \frac{\mu}{4\pi} e^\gamma (1 - H^2 / 6\mu^2) \, ,
\ee
where we expand the exponent for small $H/\mu$, which shows clearly how in this simple case the first correction that appears is quadratic in $H$ rather than linear in $H\sim L^{-1}$ as we found above~(\ref{linear2}) for the torus case.

\subsection{Vacuum energy}
The topological susceptibility we have just calculated in eqs.~(\ref{linear2}) and~(\ref{topCurv}) bears a close connection with the vacuum energy of the system.  Indeed it is well known~\cite{Wvv} that, in 4d, one can express the topological susceptibility as a second derivative of the zero-point energy with respect to the $\theta$ angle of QCD (notice the minus sign):
\be
\label{vacuum}
\chi(0) = - \left. \frac{\partial^2\rho_{\mathrm{vac}}(\theta)}{\partial \theta^2} \right|_{\theta=0} \, .
\ee

This relation will turn out to be of fundamental importance when translating the results obtained in this paper to the real 4d world, because it shows how a constant linear correction $1/L$ to the topological susceptibility, in the case of a torus, automatically filters in the vacuum energy.  As we saturated the topological susceptibility with the KS ghost, one can interpret this result as the appearance of the linear correction $1/L$ in the corresponding ghost's matrix elements. In such a formulation it has a direct analogy with the 4d model where the Veneziano ghost would return the same answer~\cite{4d}. 

The relation (\ref{vacuum}) is of uttermost relevance because, given that one knows the $\theta$ dependence of the vacuum energy~\cite{Wsp,DiV}, one can directly compute the energy mismatch that would arise between a theory in the ideal Minkowski spacetime and the actual spacetime of size $H^{-1}$ describing the Universe.  According to the general principles stated in the introduction this energy mismatch in this framework is interpreted as the observed cosmological constant $\rho_{\Lambda}$.

The vacuum energy of a QFT system is typically associated with the highest momenta fluctuations of the field in a given spacetime, where the physics at the very large distances $\sim L$  has no say.  However, as it is shown in eq.~(\ref{vacuum}), the actual vacuum energy could be related to the topological charge of the system, which, first of all, is a topological quantity and as such linked to the \emph{global} features of the spacetime, and secondly, in some cases receives corrections that are ascribable to the non-local IR properties of the theory, \emph{not} with the short distance UV.

Comfortingly, if we were now to write down the stress tensor for the bosonic version of 2d QED on a curved space, we would find that the coefficient of the cosine potential is still a constant, but exhibits a small correction that corresponds to the correction~(\ref{topCurv})--or eq.~(\ref{linear2}) for a flat torus.  This energy momentum tensor is now conserved as it must be.

\section{Conclusion}

In this work we have addressed the problem of calculating the vacuum energy in an arbitrary 2d spacetime using the KS ghost saturation for the topological susceptibility within the framework of 2d QED. It allows to restore the $\theta$-dependent portion of the vacuum energy itself. At the same time, the $\theta$-independent contribution to the vacuum energy is not linked to the ghost, and therefore it is not sensitive to the parameters of the manifold such as $L$ if it is much larger than any other scales of the problem. In this case, the cosmological constant, which is defined in our framework as the difference between the vacuum energy computed on a non-trivial manifold and flat Minkowski space,
will be entirely determined by the $\theta$-dependent portion of the vacuum energy which is the subject of the present paper.

In particular, we computed explicitely the dependence of the vacuum energy on the parameters that define the manifold we are working with, such as curvature, Hubble size, linear size $L$, etc\ldots in 2d QED. Ultimately we are interested in extending these results to the more relevant and actual case of 4d QCD~\cite{4d}, where no analytical exact results similar to those discussed above exist.  However, a close analogue of the most important element, the ghost, which is responsible for all the crucial results in 2d QED, is also present in 4d QCD. Hence, it is very naturally expect that the linear corrections we found in 2d QED will be also present in 4d QCD.
 
The result~(\ref{linear2}) is central in our work: it shows how the \emph{linear} power of the size of the torus enters the expression for the topological susceptibility, and therefore how the first correction to its Minkowski value is, barring degeneracies among the torus sides, proportional to $1/\mu L$ where $\mu$ is the ``photon mass'' in the Schwinger model, the equivalent of the $\eta'$ mass in the 4d realisation of this mechanism/system as understood by Veneziano and Witten.

The appearance of the linear (in the torus size) correction to the topological susceptibility is intimately linked with the existence of the 2d KS ghost's pole, for without it it would be impossible that a local massive scalar theory carries information about the boundaries of the manifold onto which it is quantised.  The masslessness of the ghost's pole, which is unobservable in gauge invariant quantities but has a deep impact on observables related to gauge-variant operators, is a key feature of this model and necessary ingredient for this result; the ghost's pole is the carrier of the long-distance information stored in the boundary conditions of the system.  Notice that the ghost's pole is \emph{not} lifted by interactions with quarks; however, the corresponding matrix elements are slightly different from their Minkowski values, which eventually leads to the preeminent result (\ref{linear2}). 

The second crucial observation to be made about the linear term is that its existence owes to the presence of the mass term for the fermion, without which the topological susceptibility automatically vanishes in every spacetime, as ensured by the WI.  This says that as long as the system comprises only massless fermions, the linear corrections to the chiral condensate~(\ref{condTexpl}), although still there, will not be observable.

At last, let us conclude with a few comments on the relevance of these results for the 2d and 4d vacuum energy.  The linear correction found in equation~(\ref{linear2}) leads directly to a corresponding correction in the vacuum energy of the system, as seen in~(\ref{vacuum}).  This means that the physical vacuum energy is not entirely determined by internal  properties of the system under investigation, but, in some cases, it may depend on the large distances through the macroscopic parameters that characterise the manifold, e.g.\ the linear length $L$. It is very different from the simple expectation that the physics must not be sensitive to very large distances, and that boundary effects must  vanish as $\exp{(-\mu L)}$.

Not only the relevant parameters describing the vacuum energy are found to be in the IR sector of the quantum theory, they are the direct expression of the \emph{global} topological properties of the embedding spacetime.  Therefore, in this framework, vacuum energy (the cosmological constant in 4d) is a consequence of a non-trivial global topology, whose macroscopic properties are carried by the massless KS ghost's pole in 2d.  Note how this argument can be generalised immediately to 4d, although in that case very little can be analytically computed, but some numerical calculations on the lattice can be done as suggested in \cite{4d}.

The first correction to the vacuum energy thereby identified is hence found to be proportional to the linear correction $1/\mu L$, which can be rewritten as $H/\mu$ where the Hubble parameter is taken as the size of the observable Universe today.  In 2d then, the remnant vacuum energy is expressed as $\rho_{\mathrm{vac}} \simeq mH$; in 4d one would instead find~\cite{4d}:
\be
  \label{final}
  \rho_{\Lambda}\sim 
  c\cdot\frac{2H}{m_{\eta'}}\cdot  |m_q\la\bar{q}q\ra  | \sim c(3.6\cdot 10^{-3} \text{eV})^4, 
  \ee
with $c\sim 1$ is a coefficient of order one. This estimate is to be compared with the observational value $ \rho_{\Lambda}= (2.3\cdot 10^{-3} \text{eV})^4$.

The similarity in magnitude between these two values is very encouraging. In our view, it is a clear indication that the cosmological vacuum energy, confinement, $U(1)$ problem, and topology are intertwined and their interrelations need to be explored further, as they could lead us to the solution of this intricated cosmological puzzle.

Let us also point out that the result~(\ref{final}) is based on our understanding of the ghost's dynamics: it can be analytically computed in the 2d Schwinger model and hopefully it can be tested in 4d QCD using lattice QCD computations.  This contribution to the vacuum energy is computed using QFT techniques in a static non-expanding universe.  As it stands, it can not be used for studying its evolution with the expansion of the universe.  In order to do so one needs to know the dynamics of the ghost field coupled to gravity on a finite manifold.

A final comment on our definition (or prescription) for the physical vacuum energy.  As we have discussed in the introductory sections, we define the observable vacuum energy as the differential stress tensor between infinite Minkowski and finite compact spacetime.  Therefore, with this prescription, all the usual contributions such as gluon condensates, or the condensate from the Higgs field, etc., will cancel out in the subtraction as they appear with almost equal magnitude in both compact size $L$ and non-compact manifolds. The relevant difference will behave as $\exp(-m L)$ due to their massiveness and can be safely neglected.   The Veneziano ghost's contribution is unique in all respects: its masslessness is protected and is therefore the only field linearly sensible to the global topology.

If the existence of the linear $1/L$ correction is confirmed by 4d lattice calculations, it may have profound and far reaching consequences, some of which (in particular in cosmological context, directly related to this work) can be tested in present and future Cosmic Microwave Background experiments as suggested in~\cite{Urban:2009ke}.

\section*{Acknowledgements}
FU thanks R Woodard for useful conversations.  This research was supported in part by the Natural Sciences and Engineering Research Council of Canada.

\section*{Appendix A - Conventions}\label{conv}
We are working in 2d curved spacetime.  The indices conventions are as follows:
\be
a,b,\ldots = \mathrm{Flat \,\, space} \qquad \mu,\nu,\ldots = \mathrm{Curved \,\, space} \nonumber
\ee
The 2d Minkowski metric is given by
\be
\eta_{ab} = \mathrm{diag}(1,-1) \, .
\ee
The 2d Minkowski matrices can be chosen as:
\be
\left\{
\begin{array}{l}
\gamma^0 = \sigma_2 = \left( 
\begin{array}{cc}
0 & -i \\
i & 0 \\
\end{array} \right) \\
\\
\gamma^1 = i \sigma_1 = \left( 
\begin{array}{cc}
0 & i \\
i & 0 \\
\end{array} \right)
\end{array} \right. \, ,
\ee
and they satisfy
\be
\left\{ \gamma^a , \gamma^b \right\} = 2\eta^{ab} \, .
\ee

The axial matrix is defined as
\be
\gamma^5 = \gamma^0 \gamma^1 = \sigma_3 = \left( 
\begin{array}{cc}
1 & 0 \\
0 & -1 \\
\end{array} \right) \, ,
\ee
which gives the following rule in 2d
\be
\gamma_a \gamma^5 = \epsilon_{ab} \gamma^b \, ,
\ee
with
\be
\epsilon_{ab} = \left( 
\begin{array}{cc}
0 & 1 \\
-1 & 0 \\
\end{array} \right) \, .
\ee

Now let us define the curved 2d spacetime.  As any 2d spacetime it is conformally flat, hence, in standard notation,
\be
g_{\mu\nu} = \Omega(x)^2 \eta_{\mu\nu} \quad \Rar \quad g = \mathrm{det} \, g_{\mu\nu} = \Omega^4 \, ,
\ee
and
\be
\left\{
\begin{array}{l}
g_{\mu\nu} = e^a_\mu e^b_\nu \eta_{ab} \\
g^{\mu\nu} = e_a^\mu e_b^\nu \eta^{ab}
\end{array} \right.
\quad \Rar \quad \left\{
\begin{array}{l}
e^a_\mu = \Omega \delta^a_\mu \\
e_a^\mu = \Omega^{-1} \delta_a^\mu
\end{array} \right. \, .
\ee
The curved $\gamma$ matrices are defined through
\be
\left\{
\begin{array}{l}
\gamma^\mu = e^\mu_a \gamma^a \\
\gamma_\mu = e^a_\mu \gamma_a
\end{array} \right. \, .
\ee
Finally, the antisymmetric $\epsilon$ becomes
\be
\left\{
\begin{array}{l}
\epsilon_{\mu\nu} = \delta^a_\mu \delta^b_\nu \epsilon_{ab} \\
\epsilon^{\mu\nu} = \delta_a^\mu \delta_b^\nu \epsilon^{ab} / g
\end{array} \right.
\quad \Rar \quad \left\{
\begin{array}{l}
\epsilon_{\mu\lambda} \epsilon^{\lambda\nu} = \delta^\nu_\mu / g \\
\epsilon_{ac} \epsilon^{cb} = \delta_a^b
\end{array} \right. \, .
\ee

The spinor covariant (in the GR sense) derivative operator in the 2d curved QED lagrangian is defined as
\be
\slash D = \gamma^\mu D_\mu = \gamma^\mu \stackrel{\leftrightarrow}{\partial_\mu} + \gamma^\mu \omega_\mu\, ,
\ee
with $\stackrel{\leftrightarrow}{\partial_\mu}$ acting only on the spinors, and $\omega = 0$ in 2d;  the vector covariant derivative is instead
\be
D_\mu = \partial_\mu + \Gamma_\mu \, ,
\ee
so that the curved and flat space box operators are
\be
\Box = g^{\mu\nu} D_\mu D_\nu \quad \mathrm{and} \quad \bar{\bar\Box} = \eta^{ab} \partial_a \partial_b \, .
\ee

In section~\ref{curv} we also need a compact 2d spacetime with euclidean signature of length $L$.  This can be parametrised in the most general way using quasi-isothermal coordinates as
\be\label{coords}
g_{\mu\nu} = e^{2\sigma} \left(
\begin{array}{cc}
|\tau|^2 & \tau_1 \\
\tau_1 & 1 \\
\end{array} \right) \, ,
\ee
where $\tau = \tau_1 + i \tau_0$ is the Teichm\"uller parameter on the 2-torus, and $\sigma(x) = \ln\Omega(x)$ is the gravitational Liouville field.  When studying applications to finite temperature field theory the identification $\tau = i\beta/L$ holds.

\section*{Appendix B - The WI and the $\eta'$ mass}\label{WIapp}
We want to derive the relevant WI appearing in the body of this paper.  One method one can employ is to perform a chiral rotation $\psi \rar \exp(i\gamma_5\alpha)$ with $\alpha(x)$ an infinitesimal (fictiously gauged) parameter, and then demand that the generating functional for the connected part of the action, customarily defined as $W$ does not change.  Let us begin with the lagrangian
\be
{\cal L} = {\cal L}_{\mathrm{QED}_2} + \theta Q + S \Phi + S_5 \Phi_5 \, ,
\ee
where
\be
Q = \frac{q}{2\pi} \epsilon_{ab} F^{ab} \, ,
\ee
is the topological charge density, and we have introduced sources for the $\theta$-term, and the (pseudo)scalar $\Phi = \bar\psi \psi$, $\Phi_5 = \bar\psi \gamma_5 \psi$.  In principle there will be sources for the vector and axial currents $j_a = \bar\psi \gamma_a \psi$ and $j_a^5 = \bar\psi \gamma_a \gamma^5 \psi$, as well as for the fermion and gauge boson, but they are not relevant in this derivation.

The effect of the infinitesimal chiral rotation $\psi \rar \exp(i\gamma_5\alpha)$ is to shift the lagrangian (at $\theta = 0$) by the amount
\be\label{deltaL}
\delta{\cal L} = \alpha \left[ \partial^a j^5_a - 2im \Phi_5 + Q + 2i S \Phi_5 + 2i S_5 \Phi \right] = 0 \, , \nonumber\\
\ee
where the factor $Q$ is due to the Jacobian of the transformation (notice the factor of 2 compared to the finite transformation Jacobian in eq.~(\ref{jac}).  One can rewrite this equation in terms of functional derivatives of the effective action $W = - i \ln Z$ as
\be
\partial^a \frac{\delta W}{\delta V_5^a} - 2i m \frac{\delta W}{\delta S_5} + \frac{\delta W}{\delta \theta} + 2i S  \frac{\delta W}{\delta S_5} +2i S_5  \frac{\delta W}{\delta S} = 0 \, , \nonumber\\
\ee
which can be differentiated again with respect to $\theta$ or $S_5$, and the resulting equations combined provide the crucial zero momentum WI
\be\label{wiW}
\frac{1}{4}  \frac{\delta^2 W}{\delta \theta^2} = m \frac{\delta W}{\delta S} - m^2 \frac{\delta^2 W}{\delta S_5^2} \, ,
\ee
where all sources have been turned off.  This can be written in a more familiar form as
\be
\frac{i}{4} \int \dd^2x \left< T_W Q(x) Q(0) \right> = m \left<\bar\psi \psi \right>_{\theta=0} + {\cal O}(m^2) \, ,
\ee
which is formula~(\ref{WI}) in the paper. In this WI we have explicitely appended a subscript $W$ in the T-product, standing for Wick T-product which includes the contribution from unphysical states, as opposed to the usual Dyson T-product in which only physical states contribute.  The two definitions in coordinate space differ by a delta function, which is essential if one has to satisfy the WI (\ref{WI}), as will be shown explicitely in the chiral limit.

As mentioned previously, the topological susceptibility (see~(\ref{topDef})) is intimately linked to the celebrated Veneziano proposal for the $\eta'$ problem, which was inspired by the use of a ghost's pole in the 2d Schwinger model as pushed forward by Kogut and Susskind.  In short, the problem, and its solution, can be formulated as follows.  4d QCD possesses an approximate SU(3)$\times$SU(3) chiral symmetry which is spontaneously broken by the chiral condensate, and as such it generates light pseudoscalar mesons which are the corresponding almost Goldstone bosons.  The dynamics of the octet formed by the $\pi$, the $K$ and the $\eta$ is well described in terms of this approximation scheme, including their masses which are light because of the soft explicit violation of the chiral symmetry.  However, just as for the octet, also the ninth meson $\eta'$ should be almost massless (exactly massless in the chiral limit).  Yet, the $\eta'$ meson is heavy~\cite{pdg} and it is not possible to ascribe its mass to the light $u$, $d$, and $s$ masses.

One may observe that the axial current is actually anomalous, and therefore even in the chiral limit, will not demand the $\eta'$ mass to vanish.  However, this observation, albeit true (and essential in what follows), is not sufficient to explain the large value of the ninth meson mass.  One instead needs a non-zero topological susceptibility of the vacuum, as it can be seen from the famous Witten-Veneziano relation~\cite{Wvv,Ven} (another by-product of the WI), where the $\eta'$ mass is related to the square root of the condensate term
\be
m_{\eta'}^2 = \mu^2 = \frac{4}{|f_{\eta'}|^2} \chi(0)^{\mathrm{gauge}} \, ,
\ee
where, for the Schwinger model, the ``$\eta'$'' decay constant is given by $i/\sqrt\pi$, and the apex ``gauge'' means calculated in the pure gauge theory.

A comment on renormalisation and the WI. It is possible to show, using the WI's, that the topological susceptibility at zero momentum is a Renormalisation Group Equation (RGE) invariant.  This comes about straighforwardly from the fact that the WI's we have derived are valid for bare or renormalised quantities alike; this implies that, for instance, the conserved and gauge invariant vector current is not renormalised, and that the anomalous axial current WI eq.~(\ref{deltaL}) is RGE invariant, as it should.  Use of these properties lead to the following RGE for the topological susceptibility
\be\label{rge}
{\mathscr D} \chi(0) = 0 \, ,
\ee
where ${\mathscr D}$ is the appropriate RG derivative operator, see~\cite{Shore90,Shore91}.

%\appendix

%\begin{acknowledgments}
%\end{acknowledgments}

\end{document}